\documentclass[12pt]{report}

\usepackage{graphicx}
\usepackage{float}

\usepackage{amssymb}
\usepackage{multirow}
\usepackage{multicol}
\usepackage[square,numbers]{natbib}
\usepackage{longtable}
\usepackage{fancyhdr}
\usepackage{tikz}
\usepackage{pdfpages}
\usepackage{adjustbox}

\usepackage[framemethod=TikZ]{mdframed}
\mdfsetup{
    roundcorner=6pt,
    linewidth=1.5pt,
    frametitlerule=true,
}

\usepackage{dirtytalk}

\usepackage{color, soul, colortbl}

\usepackage{titlesec} 
\usepackage[left=4cm,right=4cm,top=2.5cm,bottom=4cm]{geometry}

\usepackage{tabularx}

\definecolor{DarkRed}{RGB}{163,22,31}
\definecolor{LightRed}{RGB}{247,14,45}
\definecolor{PaleRed}{RGB}{239,134,148}
\definecolor{cpiGray}{RGB}{106,100,100}

\titleformat*{\section}{\color{DarkRed}\normalfont\bfseries\LARGE}
\titleformat*{\subsection}{\color{LightRed}\normalfont\bfseries\LARGE}
\titleformat*{\subsubsection}{\color{PaleRed}\normalfont\bfseries\large}

\def\title#1{\gdef\@title{#1}\gdef\reporttitle{#1}}

\title{Automatic  User  Profiling in Darknet Markets}
\author{Claudia Peersman, Matthew Edwards, Emma Williams \& Awais Rashid}

\renewcommand{\maketitle}{\newpage
\newgeometry{margin = 0in}
\includegraphics[width=3.09in]{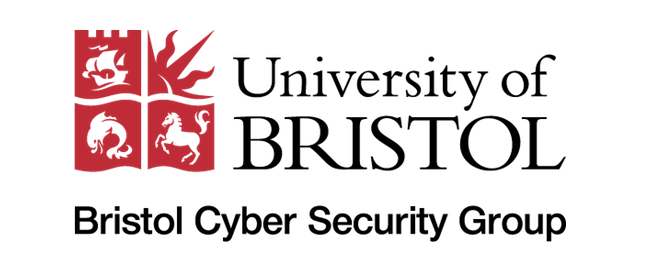}
\setlength{\fboxsep}{0pt}
\hfill \colorbox{cpiGray}{\makebox[3.22in][r]{\shortstack[r]{\vspace{2.75in}}}}%
\vspace{-0.25pt}
\setlength{\fboxsep}{10pt}
\setlength{\fboxrule}{0pt}
\colorbox{DarkRed}{\makebox[8.25in][l]{\hfill \shortstack[r]{\fontsize{36}{36}\rmfamily\color{white} \reporttitle\\%
\fontsize{24}{24}\rmfamily\color{white}}}}%
\setlength{\fboxsep}{0pt}
\vspace{-8.5pt}
\hfill \colorbox{cpiGray}{\hspace{.25in} \parbox{2.97in}{\vspace{3.6in} \color{white} \textbf{Claudia Peersman \\ Matthew Edwards \\ Emma Williams \\ Awais Rashid \\\\   \today \vspace{2.3in} \vfill}}}%
\let\Title\title
\restoregeometry}

\mdfdefinestyle{SummaryBox}{%
    linecolor=DarkRed,
    outerlinewidth=3pt,
    roundcorner=20pt,
    innertopmargin=\baselineskip,
    innerbottommargin=\baselineskip,
    innerrightmargin=20pt,
    innerleftmargin=20pt,
    frametitlerule=true,
    frametitlerulecolor=DarkRed,
    frametitlefontcolor=white,
    frametitlebackgroundcolor=DarkRed,
    fontcolor=DarkRed,
    frametitle=Summary,
    backgroundcolor=gray!50!white,
    leftmargin=18pt,
    rightmargin=18pt}

\begin{document}

\begin{titlepage}
\maketitle
\end{titlepage}
\newpage

\section*{Automatic  User  Profiling in Darknet Markets -- a Scalability Study}

To extend our qualitative analysis we reviewed a range of Natural Language Processing and Text Categorisation techniques that could be deployed on online communications in Darknet Marketplaces. Our aim was to remove the observer-observed effect, which would allow us to explore cybercriminal interactions without having to rely on the self-reporting of individual users. While recent advances in these fields have shown promising results with regard to detecting demographic characteristics, such as age group and gender, in several text genres by automatically analysing the variation of an author's linguistic features, applying such techniques on underground cybercriminal communications differs from more general applications in that its defining characteristics are both domain and process dependent. This gives rise to a number of challenges of which contemporary research has only scratched the surface. More specifically, a text mining approach applied on online communications typically has no control over the dataset size -- the number of available communications will vary across users. Hence, an automated system has to be robust towards limited data availability. Additionally, the quality of the data cannot be guaranteed. As a result, the approach needs to be tolerant to a certain degree of linguistic noise (for example, abbreviations, non-standard language use, spelling variations and errors, non-standard use of punctuation). Finally, in the context of cybercriminal fora, it has to be robust towards deceptive or \textit{adversarial} behaviour, i.e. offenders who attempt to hide their identity and criminal intentions (\textit{obfuscation}) or who assume a false digital persona (\textit{imitation}), potentially using coded language.

In this study, we investigate the scalability of state-of-the-art user profiling technologies across different online domains. More specifically, this work aims to understand the reliability and limitations of current computational stylometry approaches when these are applied to underground fora in which user populations potentially differ from other online platforms (predominantly male, younger age and greater computer use) and cyber offenders who attempt to hide their identity. Because no ground truth is available and no validated criminal data from historic investigations is available for validation purposes, we have collected new data from clearweb forums that do include user demographics and could be more closely related to underground fora in terms of user population (e.g., tech communities) than commonly used social media benchmark datasets showing a more balanced user population. 

\newpage

Additionally, a key question when designing a user profiling system that could be used to support cybercrime investigations, is if the methodology will remain useful when it is confronted with adversarial text samples. Initial work in this area is not very encouraging. The authors of \cite{brennan2012adversarial}, for example, demonstrated that even state-of-the-art authorship identification methods could be reduced to random behaviour when they are confronted with passages that include obfuscation, in which people attempt to hide their identity, or imitation, during which people try to mimic an other author’s writing style \cite{brennan2009practical}. However, like previous age prediction studies most adversarial stylometry research often included relatively large samples of texts in their experiments . As a result, it could very well be that the number of clues revealing an author's own writing style were too limited to stand out between the majority of deceptive features in these larger text samples. 

Within the context of the present study, it was our hypothesis that a user's stylistic fingerprint (or rather ``writing print'') might unwittingly flow through in (parts of) their communications. Hence, this study also investigates whether a \emph{message-based} approach, in which predictions are made on the level of the individual post and aggregated to the user level in post-processing, enables the system to identify potential clues of deception more accurately than the traditional \emph{user-based} approach that renders predictions directly on the user level. 

The resulting models are evaluated using the PAN dataset \cite{rangel2015overview} and two additional datasets collected from the clearweb (the Goodreads and Pornhub datasets). Finally, the best performing models are applied to the DNM dataset. The findings from this analysis were used to inform AMoC's final qualitative framework of cyber offender characteristics. 

\subsection*{Detecting Age and Gender Online}

\subsubsection*{Dataset}

While the DNM Corpus described below generates material suitable for unsupervised learning 
approaches, it does not provide ``ground truth'' for training and testing 
user profiling models. Hence, the performance of our user profiling techniques needed to be 
assessed through first establishing a baseline in similar non-criminal data for which ground-truth 
demographics are available. For this purpose, we used the PAN Corpus (2012 -- 2017), which contains different corpora collected from the author profiling tasks at PAN 2013, 2014 and 2015 \cite{rangel2013overview,rangel2014overview,rangel2015overview} 
and covers three online media genres (blogs, Twitter feeds and unspecified social media postings). 
All corpora contain metadata about gender and age group (18--24, 25--34, 35--49, 50--XX). An overview of the Pan dataset is provided in Table \ref{tab:pan_dataset}.

\begin{table}[h!]
  \centering
  \small
  \caption{Number of messages per age and gender group in the (English) Pan dataset.}
  \label{tab:pan_dataset}
    \begin{tabular}{ccc|c}
    \hline
    \textbf{Age Group} & \textbf{Female} & \textbf{Male} & \textbf{Total} \\
    \hline
    18-24 & 5,572  & 5,856  & 11,428 \\
    25-34 & 6,243  & 5,749  & 11,992 \\
    35-49 & 2,005  & 1,797  & 3,802 \\
    50-XX & 1,013  & 917   & 1,930 \\
    \hline
    \textbf{Total} & 14,833 & 14,319 & 29,152 \\
    \hline
    \end{tabular}
  \label{tab:addlabel}
\end{table}

\subsubsection*{Feature Selection}
\label{sec:features}
Text mining studies typically include (combinations of) lexical, character and syntactic features in their experiments. In this study, a new feature type is introduced, namely \textit{chatspeak features}. We describe the NLP procedures for extracting these features at the end of this section.

\textit{Lexical features.} Unlike some previous studies on age or gender prediction, we did not create a dictionary of hand-picked features that are likely to distinguish between different age or gender categories. Instead, we applied a data-oriented approach by extracting all token unigram features (i.e., a Bag-of-Words model), including emoticons, allowing for a more complete picture of the age and gender related linguistic variation in the data. To enable a similar data-oriented approach for content and function words individually, a list of function words is required. Although such a list in itself is limited in Standard English, when analysing online social media messages, the approach is confronted with numerous linguistic variations (e.g., ``ive'' instead of ``I've''). 
Therefore, we extended the list of standard function words with the most commonly used non-standard function words found through manual inspection of the datasets. The content word features included all other words and emoticons. Additionally, to include context information in the experiments, we extracted token bigrams. 

\textit{Character features.} As character \textit{n}-grams have been shown to be useful for tracing stylometric evidence beyond topic and genre \cite{sarawgi2011gender} and proven to be reliable when dealing with limited data \cite{peersman2018detecting,luyckx2011scalability}, we included character bi-, tri- and tetragrams in the user profiling experiments. 

\textit{Chatspeak features.} In chatspeak, paralinguistic and non-verbal cues --- that are present in spoken discourse, but absent in the formal written repertoire --- are often compensated by other linguistic features, such as emoticons, character flooding and the use of upper-case and punctuation flooding to express emphasis \cite{crystal2001language}. Hence, information on the presence of these paralinguistic features were also included in the document representations. More specifically, prior to pre-processing, the occurrence of character and punctuation flooding were extracted and represented as ``[char\_flood]'' and ``[punct\_flood]'' features, and the occurrence of non-standard capitalisation as ``[char\_upper]''. Additionally, emoticons and other combinations of characters and punctuation were represented as ``[emoji]'' features. Finally, for each token, we extracted information on its ``standardness'': a word was labelled non-standard when it was analysed as such by Language Tool\footnote{https://pypi.org/project/language-tool-python/}. We also included the rule id, category and rule issue type for each non-standard word as additional features. An example is provided in Table \ref{tab:non_standard}.

\begin{table}[htbp]
  \centering
  \scriptsize
  \caption{Language Tool output for this PAN post: ``\textit{ppl who stan her at this point are as ignorant as iggy stans i love azealias insite on racism but her mess makes it invalid}''}
  \label{tab:non_standard}
    \begin{tabular}{p{3cm}p{4.5cm}p{3cm}p{3cm}}
    \hline
    Non-Standard Language Use & Rule Id & Category & Rule Issue Type \\ \hline
    Sentence does not start with an uppercase letter, stan, iggy, stans, i, azealias, insite, no comma before `but' & UPPERCASE\_SENTENCE\_START, MORFOLOGIK\_RULE\_EN, MORFOLOGIK\_RULE\_EN, MORFOLOGIK\_RULE\_EN, I\_LOWERCASE, MORFOLOGIK\_RULE\_EN, MORFOLOGIK\_RULE\_EN, COMMA\_COMPOUND\_SENTENCE\_2 & CASING, TYPOS, TYPOS, TYPOS, TYPOS, TYPOS, TYPOS, PUNCTUATION & typographical, misspelling, misspelling, misspelling, misspelling, misspelling, misspelling, typographical \\ \hline
    \end{tabular}
  \label{tab:addlabel}
\end{table}

\subsubsection*{Corpus-based Semantic Embeddings}

Corpus-based semantic embeddings exploit statistical properties of a text to embed words in vectorial space. We employ two different approaches to generating corpus-based semantic embeddings:

\begin{itemize}
    \item For each content word we obtained vector representations from a model that was pre-trained on Twitter data (glove.twitter.27B)\footnote{https://nlp.stanford.edu/projects/glove}. Using word embeddings allows for detecting semantic similarities between words based on their distributional properties in large corpora, which could boost the performance of a user profiling model. 
    \item We also included Latent Semantic Analysis (LSA) \cite{pedregosa2011scikit} to enable a comparison with the results described in \cite{rangel2015overview}. LSA is an unsupervised Information Retrieval technique for extracting and representing the contextual-usage meaning of words in a collection of documents. 
\end{itemize}

\subsubsection*{Machine Learning}
For the purpose of this study, a range of different machine learning methods were explored. More specifically, for classification, we examined the performance of the following algorithms\footnote{The NN model was trained using the Keras \cite{chollet2015keras} interface for deep learning. Keras serves as a high-level API for TensorFlow (https://www.tensorflow.org). The other classifiers were trained using the Scikit-learn \cite{pedregosa2011scikit} machine learning package.}.

\begin{itemize}
\item \textit{Support Vector Classification}. Because SVMs have demonstrated robustness to high-dimensional data and imbalanced text mining problems (e.g., \cite{sebastiani2002machine,seiffert2014empirical}), two SVM implementations were included in the experiments: C-Support Vector Classification (with RBF kernel, $c = 2048.0$) (SVC) and a linear SVM with Stochastic Gradient Descent (SGD)\footnote{When applying SGD training, the gradient of the loss is estimated per instance and the model is updated along the way with the learning rate \cite{scikitman2}.} training. The first is standard in a text mining approach. The latter was included because of its increasing popularity in ``big data'' ML applications. 

\item \textit{Naïve Bayes (NB)}. The Naïve Bayes algorithm is popularly used for traditional text classification purposes and has shown tolerance to missing values \cite{kotsiantis2007supervised}. For classification, a Multinomial NB (gender detection) and a Complement Naive Bayes algorithm (age identification) were used, because the algorithms outperformed the Gaussian and Bernouilli NB algorithms during the preliminary experiments. The Complement Naive Bayes classifier was designed to correct the ``severe assumptions'' made by the standard Multinomial Naive Bayes classifier and was found particularly suited for imbalanced data sets \cite{scikitman2}.

\item \textit{k-Nearest Neighbor (k-NN)}. The $k$-Nearest Neighbor algorithm was suggested by \cite{luyckx2011scalability} for performing authorship attribution ``in the wild'' and has demonstrated robustness towards overfitting \cite{kotsiantis2007supervised}. The $k$ parameter was experimentally determined for each training session. Neighbor weights were assigned proportionally to the inverse distance from each test instance. 

\item \textit{Random Forest (RF)}. As the Random Forest algorithm has shown tolerance to missing values and irrelevant attributes in prior text categorisation research \cite{kotsiantis2007supervised}, which could be relevant to the task at hand, it was included in the experiments. The number of trees in the forest is set to 10; the Gini impurity function \cite{breiman1984olshen} was used to split a decision tree.

\item \textit{Ridge Classifier (RC)}. This algorithm treats a problem as a regression task and is shown to be significantly faster than, for example logistic regression algorithms, when analysing a high number of classes \cite{scikitman2}.

\item \textit{Passive Aggressive Classifier (PAC)}. The algorithm is ``passive'' whenever the loss is zero. However, even a slightly positive loss ``aggressively'' forces an update on the algorithm's hypothesis \cite{crammer2006online}. We included this algorithm in our experiments, because passive-aggressive algorithms are designed for large-scale learning and they have shown robustness to handling sparse datasets \cite{scikitman2}. 

\item \textit{Multi-Layer Perceptron (MLP)}. An MLP classifier was trained using Backpropagation. The log-loss function was optimised using stochastic gradient descent as proposed in \cite{kingma2014adam} (`adam'), which has shown efficiency towards large datasets \cite{scikitman2}. Also, MLPs have been used to derive robust features for e.g., speaker recognition \cite{heck2000robustness}.

\item \textit{Long Short Term Memory Recurrent Neural Network (NN)}. LSTM is an artificial recurrent neural network (RNN) architecture used in the field of deep learning. The authors of \cite{nowak2017lstm} have shown the superiority of this method over other algorithms for short text sentiment classification across different platforms. 
\end{itemize}

The parameters reported above were optimised during the preliminary experiments on a validation sample of the training data. 

\subsubsection*{Post-level vs. User-level Experiments}

To enable a valid comparison between the aggregated message-level and the user-level approach, we extracted all users who had produced at least fifteen posts from the PAN Corpus and created two datasets for our experiments: the post level and the user level datasets, in which the document representations per user contain 1 and all messages, respectively. None of the postings were discarded, only regrouped. 

The post level dataset was randomly divided into a 50\% training set, a 30\% aggregation set and a 20\% test set. The user level dataset was split into a 80\% training and a 20\% test set. Both test sets contained the same users and messages. During the splitting, the messages were clustered so that no user was present in two different sets. Distributing users rather than messages ensured that no user in training also appeared in the aggregation or test sets, which prevented overfitting of user-specific features. 

Evaluation of the classification system proceeded as follows:

\begin{enumerate}

\item To aggregate the predictions on the message level to the user level, the classifier produced labels for the aggregation set. The aggregation method was developed using an ensemble classifier on this output, and performance was established through five-fold cross-validation on the aggregation set. 
\item For final validation of the message-level approach, individual classifiers were trained on the training set, produced labels for the aggregation set and the test set; the ensemble model was trained on the aggregation set, and its predictions taken for the test set.
\item User-level classifiers were trained and evaluated through ten-fold cross-validation within the user level training set. The best performing classifier was then used to produce predictions for the test set, which contained data identical to the test set of the message-level experiments.
\end{enumerate}

\subsubsection*{The Message-based Approach}
\label{chap05:message-based}

Social network messages differ from other online text genres, such as e-mails or blogs, in several aspects. The length of each instance is usually much shorter, their vocabulary and grammatical structure are often non-standard and the distribution of lengths is very similar: the average post length in the Pan dataset is 11.1 tokens, with on average 97.2 posts per user. In this section, we examine the behaviour of a text mining approach to automatically identify social media users' age group and gender under the complex conditions of short social network postings containing non-standard language use.  

To evaluate different aspects of methodological design, first, we conducted a series of experiments, in which we examined the performance of three feature selection techniques (Document Frequency, Chi Square ($\chi_2$) and Mutual Information) and four feature representation methods (tf-idf, binary, absolute and $l2$-normalisation) on the message-level, using the Pan dataset in a ten-fold cross validation set-up. When analysing the effect of each aspect on the performance of the model, the other factors were kept constant to allow for a valid comparison. Next, we used the best performing model to compare the performance of lexical, character, syntactic, word embedding and chatspeak features using nine different machine learning techniques (SVC, SGD, NB, $k$-NN, RF, RC, PAC, MPL and NN) for both tasks. The results for each task are described below.

\textbf{Age Group Identification}

As can be seen in Table \ref{tab:pan_dataset}, the PAN dataset is highly skewed with regard to the number of posts available for each age group. Hence, we calculated a random baseline using a stratified random classification strategy, which generates random predictions while respecting the training set class distribution. For the age group identification task, this resulted in a random baseline f-score of 39.2\% for 18--24; 41.1 \% for 25--34; 12.6\% for 35--49; and 6.4\% for 50--XX; and an average micro f-score for this task of 34.4\%.

With regard to feature selection, almost every machine learning approach benefited from reducing its dataset's dimensionality to its 10,000 most discriminative features using the Chi Square method. Additionally, our models tended to yield slightly better results when using a  binary representation method in our preliminary experiments. Hence, these were used when setting up the full experimental regime using different feature types and machine learning algorithms.

Our Long Short Term Memory Recurrent Neural Network (NN) was trained using the Glove Twitter word embeddings model (glove.twitter.27B)\footnote{https://nlp.stanford.edu/projects/glove}. Hence, we could only produce results using word embeddings. We included the Rectified Linear Unit activation function (ReLu=32) and the Softmax activation function, along with an LSTM layer containing 128 memory units. Because age group identification is a non-binary task, categorical crossentropy was used as the loss function, and the Adam algorithm \cite{kingma2014adam} as optimizer. Finally, the model took 30 passes (epochs) over each training partition in the dataset for periodic evaluation.

As can be seen in Table \ref{tab:microF_age_detection}, the best overall performance of 59.5\% (micro f-score) significantly outperforms the random baseline for age group identification and was achieved when training the Complement Naive Bayes classifier (NB) using character \textit{n}-gram features. When analysing the results for each age group individually, we found that character \textit{n}-grams also outperformed the other feature types for 3 out of 4 age groups, but in some cases different algorithms produced better results than the NB classifier, resulting in a best f-score of 71.0\% for 18--24 (NB), 33.1\% for 35--49 (NB) and 26.3\% for 50-XX (SGD). Interestingly, we found that chatspeak features (i.e., non-standard language use and paralinguistic features, see Section \ref{sec:features}) yielded the best results for the 25--34 age group (67.7\%, NB). This is in line with the work of \cite{peersman2016effects}, who found a slight increase in the chatspeak word probability between the ages of 29 and 33 in a (Flemish) Dutch dataset of online social media messages. However, these findings differ notably with previous spoken discourse studies describing the \textit{Age Grading} principle, which states that the usage of non-standard linguistic varieties tends to peak during adolescence (15--17 years old), but as social pressure increases and the use of standard language becomes more important, (for example, for building a career or raising children), people are more inclined to adapt to society's norms. Hence, the use of standard (or prestige) forms was found to peak between the ages of 30 and 55 \cite{tagliamonte2012variationist}. This could be linked to the fact that this age group entails the first generation that acquired ``the art'' of online chatting. As a result, this could indicate that the use of chatspeak in social media communications is not only attractive for adolescents, but also --- to some extent --- for people that are currently in their early thirties, because it distinguishes them from the older age groups that did not learn to chat during adolescence. However, this hypothesis requires further research to be confirmed. We provide an overview of the f-scores per age group and machine learning method in Table \ref{tab:age1_postlevel}.

\begin{table}[h]
  \centering
  \scriptsize
  \caption{Micro F-scores (\%) per feature type and machine learning algorithm for age group detection.}
  \label{tab:microF_age_detection}
    \begin{tabular}{p{3cm}ccccccccc}
    \hline
    \textbf{Features} & \textbf{SVC} & \textbf{KNN} & \textbf{RF} & \textbf{NB} & \textbf{RC} & \textbf{SGD} & \textbf{PAC} & \textbf{MPL} & \textbf{DL} \\ \hline
    BOW   & 54.3  & 40.6  & 53.6  & 56.9  & 56.6  & 57.8  & 53.6  & 54.7  & / \\
    Content Words & 53.9  & 39.4  & 53.5  & 56.6  & 56.1  & 53.7  & 53.3  & 53.5  & / \\
    bigrams & 41.6  & 28.2  & 48.8  & 50.3  & 48.4  & 47.6  & 46.9  & 50.2  & / \\
    char n-grams & 56.5  & 45.9  & 55.7  & \textbf{59.5}  & 53.7  & 58.3  & 56.7  & 58.9  & / \\
    Chatspeak Feats. & 43.4 & 44.6 & 46.4 & 40.6 & 43.6 & 43.6 & 39.6 & 45.8 & / \\
    Word Embeddings & 54.0  & 52.8  & 54.3  & /     & 52.6  & 53.1  & 48.8  & 53.8  & 48.7 \\
    LSA   & 50.1  & 47.3  & 50.1  & /     & 48.6  & 48.6  & 44.4  & 51.4  & / \\ \hline
    \end{tabular}
\end{table}

\begin{table}[h]
\scriptsize
\label{tab:age1_postlevel}
\caption{F-scores (\%) per age group and machine learning model on the post level.}
\begin{tabular}{|p{1cm}|p{2.5cm}|c|c|c|c|c|c|c|c|c|}
\hline
\textbf{Age Cat.} & \textbf{Features} & \textbf{SVC}  & \textbf{KNN} & \textbf{RF} & \textbf{NB}  & \textbf{RC} & \textbf{SGD}  & \textbf{PAC} & \textbf{MPL} & \textbf{DL} \\ \hline
\multirow{7}{*}{18--24}                & BoW               & 68.1          & 56.3         & 65.4        & 69.0          & 67.5        & 68.8          & 62.8         & 64.2         & /             \\
                      & Content Words     & 67.6          & 56.1         & 63.4        & 68.0          & 67.1        & 68.1          & 63.9         & 63.2         & /             \\
                      & Bigrams           & 60.4          & 57.2         & 53.4        & 60.2          & 56.5        & 59.2          & 59.7         & 56.4         & /             \\
                      & Char. n-grams     & 70.7 & 56.9         & 69.9        & \textbf{71.0}          & 63.5        & 68.6          & 66.3         & 69.0         & /             \\
                      & Chatspeak Feats.  & 30.6          & 47.8         & 39.3        & 20.8          & 32.0        & 31.5          & 30.7         & 36.0         & /             \\
                      & Word embeddings   & 69.8          & 64.6         & 68.3        & /             & 67.8        & 67.5          & 61.8         & 64.6         & 61.7          \\
                      & LSA               & 62.0          & 58.3         & 62.6        & /             & 59.7        & 61.2          & 55.7         & 61.5         & /             \\ \hline
\multirow{7}{*}{25--34}                 & BoW               & 61.3          & 39.6         & 60.5        & 62.8          & 60.2        & 61.5          & 57.6         & 58.8         & /             \\
                      & Content Words     & 60.4          & 36.0         & 59.6        & 62.4          & 59.7        & 60.3          & 56.2         & 57.8         & /             \\
                      & Bigrams           & 40.1          & 11.3         & 60.5        & 57.8          & 55.4        & 51.1          & 48.1         & 58.9         & /             \\
                      & Char. n-grams     & 63.7          & 50.1         & 63.8        & 62.7          & 57.7        & 61.9          & 60.9         & 63.5         & /             \\
                      & Chatspeak Feats.  & 67.2          & 56.6         & 65.0        & \textbf{67.7} & 67.0        & 66.9          & 59.3         & 66.9         & /             \\
                      & Word embeddings   & 63.5          & 55.7         & 62.7        & /             & 61.3        & 61.8          & 50.4         & 58.3         & 55.0          \\
                      & LSA               & 59.8          & 52.7         & 57.8        & /             & 59.7        & 58.1          & 49.7         & 58.3         & /             \\ \hline
\multirow{7}{*}{35--49}                & BoW               & 14.1          & 10.9         & 16.6        & 25.2          & 30.0        & 31.1          & 29.0         & 29.3         & /             \\
                      & Content Words     & 14.6          & 13.9         & 23.4        & 26.1          & 29.5        & 29.8          & 27.5         & 27.7         & /             \\
                      & Bigrams           & 8.0           & 4.4          & 15.6        & 18.4          & 19.3        & 19.2          & 20.6         & 21.9         & /             \\
                      & Char. n-grams     & 14.8          & 15.3         & 9.2         & \textbf{33.1} & 26.4        & 31.0          & 30.1         & 30.8         & /             \\
                      & Chatspeak Feats.  & 7.9           & 14.5         & 13.6        & 6.0           & 7.6         & 9.0           & 7.9          & 11.4         & /             \\
                      & Word embeddings   & 2.1           & 27.2         & 9.7         & /             & 4.9         & 7.5           & 26.0         & 27.3         & 12.8          \\
                      & LSA               & 6.0           & 17.0         & 9.0         & /             & 2.9         & 3.6           & 15.3         & 20.3         & /             \\ \hline
\multirow{7}{*}{50--XX}                 & BoW               & 6.2           & 10.6         & 12.3        & 9.3           & 20.5        & 20.5          & 21.5         & 22.1         & /             \\
                      & Content Words     & 7.8           & 10.5         & 15.4        & 11.9          & 19.9        & 19.9          & 21.9         & 19.6         & /             \\
                      & Bigrams           & 5.1           & 7.1          & 11.7        & 6.9           & 12.3        & 11.3          & 14.0         & 14.5         & /             \\
                      & Char. n-grams     & 7.8           & 11.2         & 9.8         & 22.5         & 23.4        & \textbf{26.3} & 24.3         & 24.9         & /             \\
                      & Chatspeak Feats.  & 0.3           & 13.3         & 12.9        & 0.0           & 0.0         & 0.0           & 4.1          & 8.9          & /             \\
                      & Word embeddings   & 1.7           & 14.3         & 4.9         & /             & 0.1         & 0.3           & 4.9          & 12.9         & 0.8           \\
                      & LSA               & 3.5           & 6.9          & 6.5         & /             & 2.0         & 1.3           & 1.2          & 8.8          & /            \\ \hline
\end{tabular}
\end{table}

\textbf{Gender Detection}

We performed a similar series of experiments to investigate the effect of methodological design on a text mining approach when distinguishing between male and female users in the PAN dataset, which is almost balanced with regard to the number of messages per gender category. Hence, for the gender detection task, we calculated the following random baseline scores: an F-score of 49.0\% for male and 50.9\% for female; and, as can be expected, a micro F-score of 50.0\% for this task.

Our deep learning architecture (NN) was again trained using the Glove Twitter word embeddings model (glove.twitter.27B) \footnote{http://www.aclweb.org/anthology/D14-1162}. We also included a 1D convolution layer with Rectified Linear Unit activation function (with 32 filters and kernel size = 3) and a layer with the Sigmoid activation function. Because we approached gender identification as a binary task, binary cross-entropy was used as the loss function, and the Adam algorithm \cite{kingma2014adam} as optimizer. Finally, the model also took 30 passes (epochs) over each training partition in the dataset for periodic evaluation.

As can be seen in Table \ref{tab:microF_gender_detection}, the results from the gender classification experiments are similar to those of the age prediction models: the best micro f-score on the post level of 63.8\% again significantly outperforms the random baseline for gender detection was achieved when training the NB classifier (Multinomial Naïve Bayes) using character \textit{n}-gram features. When analysing the results for the female and male categories individually, we found that BoW features outperformed all other feature types when identifying females, which resulted in a 66.9\% f-score using NB. For the male class, the best result of 66.1\% was achieved when training the $k$-NN algorithm on bigram features. We provide an overview of the f-scores per gender category and machine learning method in Table \ref{tab:results_gender_postlevel}.

\begin{table}[h]
  \centering
  \scriptsize
  \caption{Micro F-scores (\%) per feature type and machine learning algorithm for gender detection.}
  \label{tab:microF_gender_detection}
    \begin{tabular}{p{3cm}ccccccccc}
    \hline
    \textbf{Features} & \textbf{SVC} & \textbf{KNN} & \textbf{RF} & \textbf{NB} & \textbf{RC} & \textbf{SGD} & \textbf{PAC} & \textbf{MPL} & \textbf{NN} \\ \hline
    BoW   & 61.7  & 53.3  & 60    & 62.4  & 61.3  & 61.9  & 60.5  & 60.8  & / \\
    Content Words & 61    & 54.7  & 60.5  & 62.1  & 60.7  & 61.2  & 59    & 60.6  & / \\
    Bigrams & 53    & 40.2  & 54    & 56.1  & 55.4  & 56    & 56.3  & 56.3  & / \\
    Char n-grams & 63.2  & 55.5  & 62.5  & \textbf{63.8} & 60.1  & 62.9  & 62.2  &  63.7  & / \\
    Word Embeddings & 62.1  & 60    & 59.9  & /     & 60.2  & 58.7  & 52.7  & 61.1  & 60.4 \\
    LSA   & 56.9  & 54.5  & 56.3  & /     & 55.8  & 53.7  & 50.2  & 55.9  & / \\ \hline
    \end{tabular}
\end{table}

\begin{table}[h]
  \centering
  \scriptsize
  \label{tab:results_gender_postlevel}
  \caption{F-scores (\%) per gender group and machine learning model on the post level.}
    \begin{tabular}{|l|l|c|c|c|c|c|c|c|c|c|}
    \hline
    \textbf{Gender} & \textbf{Feature Types} & \textbf{SVC} & \textbf{KNN} & \textbf{RF} & \textbf{NB} & \textbf{RC} & \textbf{SGD} & \textbf{PAC} & \textbf{MPL} & \textbf{NN} \\ \hline
   \multirow{7}{*}{Male} & BoW   & 61.1  & 46.8  & 56.1  & 57.6  & 59.1  & 60.4  & 60.5  & 59.2  & / \\
          & Content Words & 58.8  & 56.9  & 57.3  & 57.8  & 58.0  & 58.2  & 57.0  & 59.3  & / \\
          & Bigrams & 50.1  & \textbf{66.1} & 46.3  & 48.8  & 52.8  & 56.7  & 57.6  & 50.9  & / \\
          & Char. n-grams & 62.3  & 52.5  & 60.2  & 60.9  & 59.4  & 62.4  & 61.8  &  62.8   & / \\
          & Chatspeak Feats. & 60.6  & 59.3  & 59.9  & 63.8  & 61.6  & 61.3  & 45.0  & 60.3  & / \\
          & Word Embeddings & 59.6  & 58.8  & 58.2  & /     & 58.3  & 55.6  & 53.5  & 61.1  & 60.6 \\
          & LSA   & 59.1  & 55.2  & 53.0  & /     & 56.9  & 57.0  & 40.3  & 56.9  & / \\ \hline
    \multirow{7}{*}{Female} & BoW   & 62.2  & 59.6  & 63.7  & \textbf{66.9} & 63.4  & 63.3  & 60.5  & 62.4  & / \\
          & Content Words & 63.1  & 52.6  & 63.6  & 66.1  & 63.4  & 64.1  & 61.0  & 61.9  & / \\
          & Bigrams & 55.8  & 15.1  & 61.5  & 63.1  & 57.9  & 55.2  & 55.1  & 61.6  & / \\
          & Char. n-grams & 64.0  & 58.4  & 64.7  & 66.6  & 60.8  & 63.4  & 62.6  &   64.6    & / \\
          & Chatspeak Feats. & 50.8  & 52.5  & 53.5  & 41.6  & 50.9  & 48.2  & 58.9  & 52.1  & / \\
          & Word Embeddings & 64.4  & 61.1  & 61.6  & /     & 62.2  & 61.7  & 51.8  & 61.0  & 60.2 \\
          & LSA   & 54.8  & 53.8  & 59.5  & /     & 54.8  & 50.4  & 59.8  & 55.0  & / \\ \hline
    \end{tabular}
\end{table}

\textbf{Boosting Strategies}

Based on the results of the systematic study of different aspects of methodological design presented above, we examined two different strategies to boost the performance for automatic user profiling using only a single message per user: a feature union approach and a balancing strategy approach. 

Because the experiments that were based on character n-gram features yielded the best micro F-score, in the next step we combined them with other single feature types. There was only one combination that was able to slightly improve upon the original performance of the character n-gram features for all age categories: when merging character n-grams with chatspeak features, the NB classifier achieved a micro f-score of 59.8\%, a 70.8\% f-score for 18--24, 62.3\% for 25--34, 32.5\% for 35--49, and 23.8\% for 50--XX. Any other combinations of single feature types (including threefold combinations) did not produce better results. 

\begin{figure}[h!]
  \caption{Results of the boosting strategies (\% f-scores) for age group identification.}
  \label{fig:age_boosting}
  \centering
  \includegraphics[width=10cm, height=7cm]{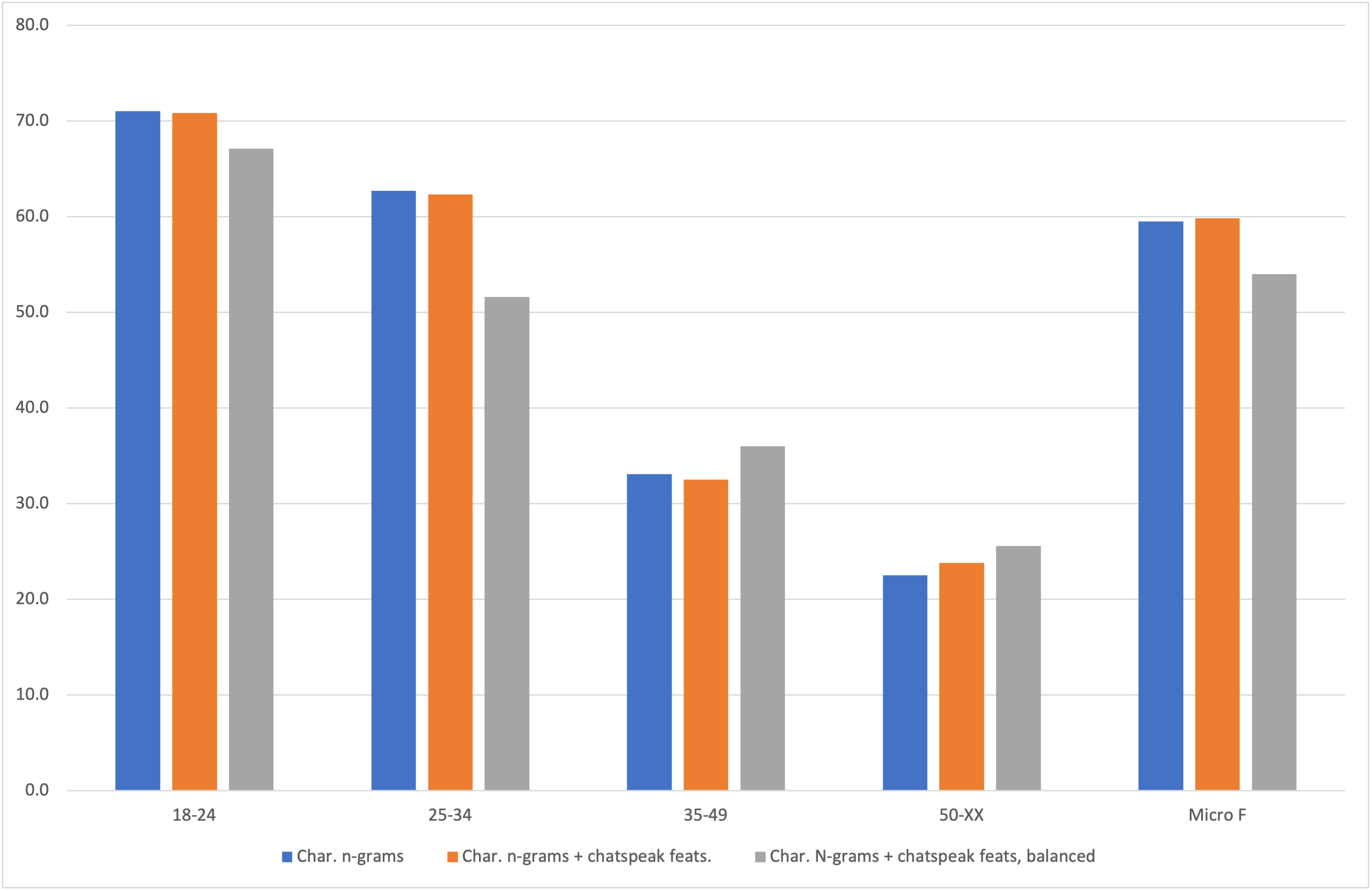}
\end{figure}

Secondly, to create a good reflection of reality, up until this point, a highly skewed data distribution was adopted during each age identification experiment. To investigate the effect of data distributions on age group identification, we balanced the data in training while maintaining the original skewed data distribution in the test partitions. We found that balancing the dataset in each training partition only, led to a considerably higher recall score for the minority classes when predicting age group compared to the imbalanced data experiments described above, but the precision decreased considerably, leading to a slightly lower micro f-score. The results are shown in Figure \ref{fig:age_boosting}.

Similar to the feature union experiments for age group detection, the best performing single feature type for gender detection was merged with other types to examine which combinations could boost the performance of the gender classifier. However, none of the combinations was able to outperform the character \textit{n}-gram model.

\textbf{Aggregating Predictions to the User Level}

For the ensemble model, we examined the performance of eight different machine learning approaches (C-SVC, $k$-NN, RF, NB, RC, SGD, PAC and MPL), using a 5-fold cross validation set-up on the aggregation data partition, to automatically aggregate the predictions on the post level to the user level. 

\begin{figure}[h!]
  \caption{Results of the ensemble learners (\% micro f-scores) for age group and gender identification on the aggregation partitions.}
  \label{fig:ensemble}
  \centering
  \includegraphics[width=10cm, height=7cm]{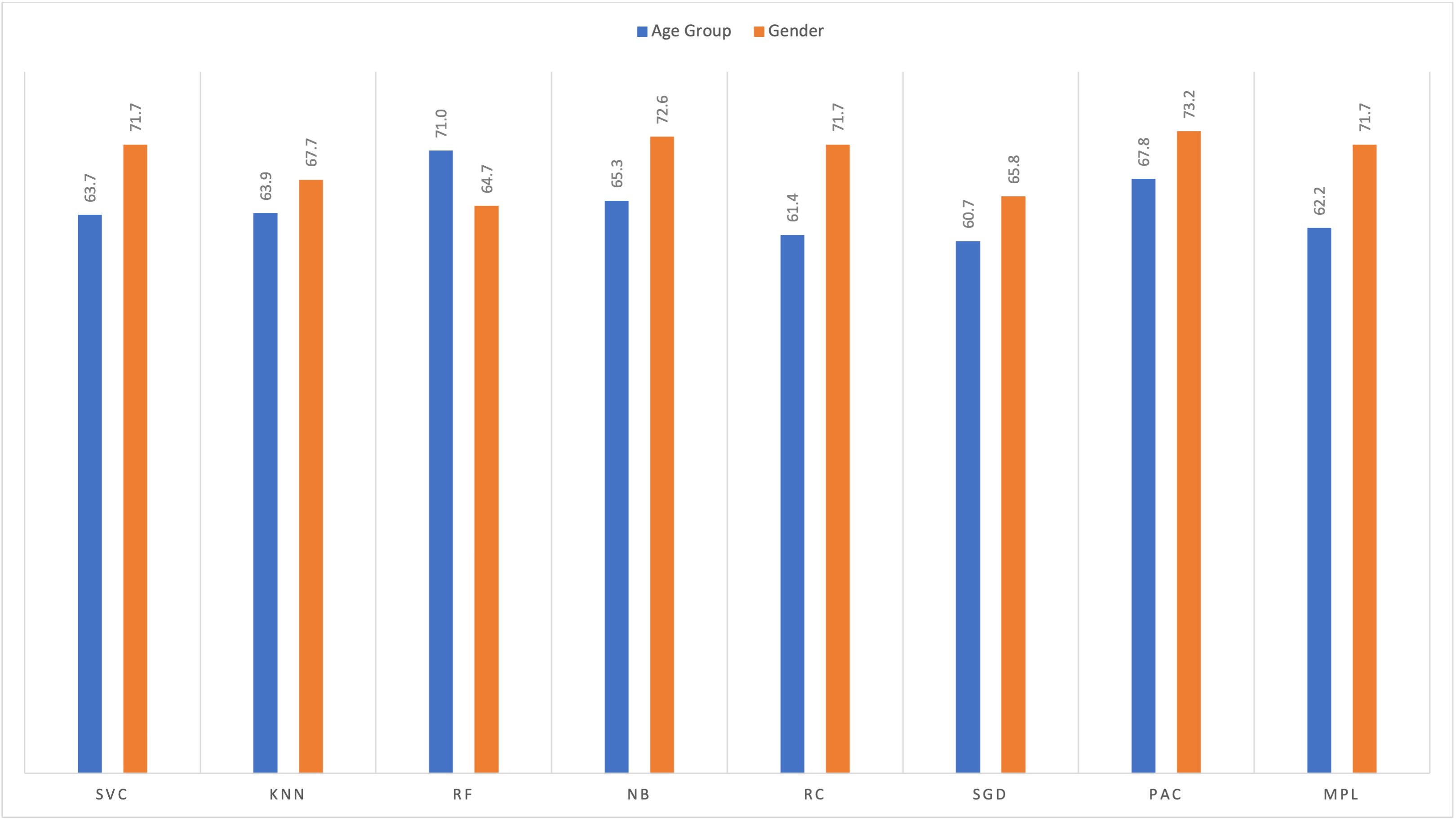}
\end{figure}

As can be seen in Figure \ref{fig:ensemble}, the performance of the age group identification model increased significantly for all learners. This time, the Random Forest (RF) outperformed the other models for both age group identification, yielding an average micro f-score of 71.0\% on the user level of the aggregation partitions. For gender detection, training the Passive Aggressive Classifier (PAC) on the output labels rendered for the aggregation set produced a best micro f-score of 73.2\%.

Finally, the best performing post level age identification model (character \textit{n}-grams + chatspeak features) was retrained on the training and aggregation partitions using the NB classifier, rendering predictions on the post level for each message in the test set. Next, the RF aggregation classifier was trained on the predictions in the aggregation partition and provided a final decision on the user level of each user in the test set. Interestingly, the performance did not drop, \textbf{resulting in a final 73.2\% micro f-score and an accuracy score of 76.7\%, an f-score of 88.9\% for 18--24; 74.5\% for 25--34; 37.5\% for 35--49 and 100.0\% for 50--XX on the user level}. 

Similarly, we used the best gender detection model (character \textit{n}-grams, NB) to produce labels on the aggregation and test partitions on the post level and trained the PAC classifier on the aggregation partition. \textbf{This resulted in a final accuracy score of 86.7\% and a micro f-score of 86.8\% for gender detection, and a final 87.5\% f-score for the female class and a 85.7\% for the male class.}

In the next section we compare our results to a user-based approach, which renders predictions directly on the user level.

\subsubsection*{The User-based Approach}
\label{chap05:user-based}
For each experiment described in this section, we collected, preprocessed and represented all messages from the same user in a single instance vector. This way, the user-based system directly labels users and no further aggregation steps are required. To enable a valid comparison with the results of the message-level experiments, we first performed 10-fold cross validation within the training partition and evaluated the best performing model on exactly the same test data as we used in the aggregated message-level experiments.

\begin{figure}[h]
  \caption{Average micro f-scores (\%) per age group and machine learning model on the user level in training.}
  \label{fig:age_user}
  \centering
  \includegraphics[width=13cm, height=10cm]{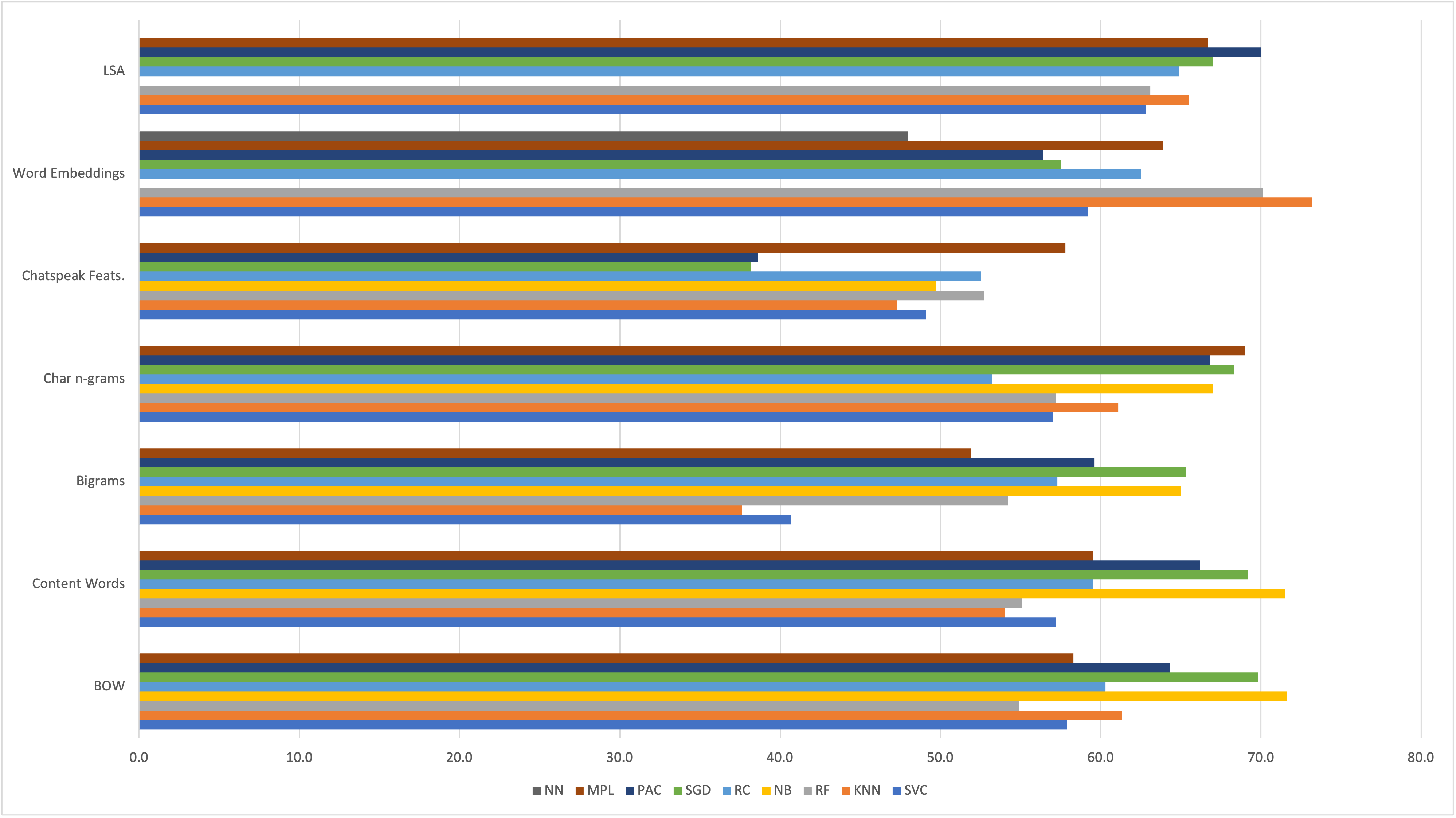}
\end{figure}

As can be seen in Figure \ref{fig:age_user}, for age group identification, the best micro f-score on the user level of 73.2\% was achieved when training a $k$-NN classifier ($k$ = 3) on word embeddings. Because no other feature combinations produced better results, this model was used to produce a final decision on the user level of each user in the test partition, which \textbf{resulted in a micro f-score of 72.0\%, an accuracy score of 73.3\%, an f-score of 76.6\% for 18--24; 76.6\% for 25--34; 52.6\% for 35--49 and 85.7\% for 50--XX on the user level}. 

\begin{figure}[h]
  \caption{Average micro f-scores (\%) per gender group and machine learning model on the user level in training.}
  \label{fig:gender_user}
  \centering
  \includegraphics[width=13cm, height=10cm]{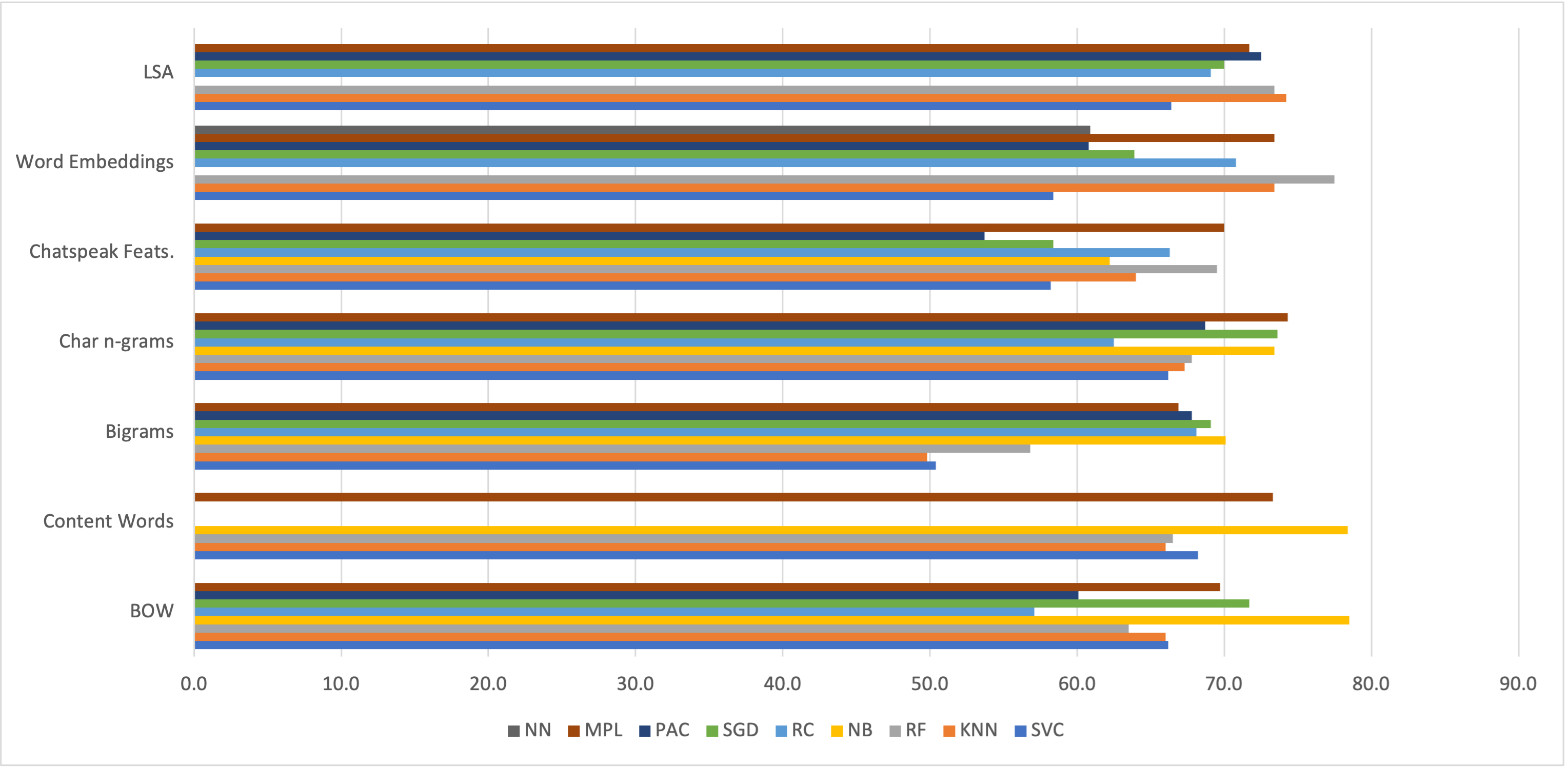}
\end{figure}

Figure \ref{fig:gender_user} shows that, with regard to gender detection, the NB classifier trained on BoW features outperformed all other feature types and learners in our user level experiments. Hence, this model was used to produce labels for each user in the test partition. \textbf{This resulted in a micro f-score of 76.5\%, an accuracy of 68.3\%, an f-score of 74.6\% for the female class and 79.2\% for the male category}.

In the next section, we set up a cross domain experiment, in which we evaluate the performance of both approaches when applied on two newly collected clearweb datasets. Finally, the best performing models are applied to the DNM dataset, which contains Darknet Market conversations between cyber offenders, in Section \ref{sec:DNM_analysis}.

\subsection*{Cross Domain User Profiling}

\subsubsection*{Datasets}

To enable an estimation of the performance of our user profiling models across different online domains, we have developed scraping techniques to collect new data from clearweb forums that (i) include user demographics (e.g., Goodreads)\footnote{https://www.goodreads.com/}, and (ii) are estimated to relate more closely to underground forums in terms of user population than the traditional social media benchmark datasets used in related work (Pornhub\footnote{https://www.pornhub.com/}). 
The general user demographics for these new clearweb forums are presented in Table~\ref{tab:user_profiles}. We display the top-5 age groups for Goodreads and Pornhub in Table~\ref{tab:age_group}. In addition, the results for top 5 countries are shown in Table~\ref{tab:top_countries}. 

\begin{table}[h!]
\scriptsize
\centering
\begin{tabular}{|c|c|c|c|c|c|}
\hline
\multirow{2}{*}{\textbf{Community}}                                     & \multirow{2}{*}{\textbf{Total users}} & \multicolumn{3}{c|}{\textbf{Gender}}              & \multirow{2}{*}{\textbf{\begin{tabular}[c]{@{}c@{}}Messages \\ (avg.)\end{tabular}}} \\ \cline{3-5}
                                                                        &                                       & \textbf{Male} & \textbf{Female} & \textbf{Others} &                                                                                      \\ \hline
\textbf{\begin{tabular}[c]{@{}c@{}}Goodreads \\ (GR)\end{tabular}}     & 100,061                               & 32\%          & \cellcolor{blue!25}67\%            & 1\%             & 26                                                                                   \\ \hline
\textbf{\begin{tabular}[c]{@{}c@{}}Pornhub \\ (PH)\end{tabular}}       & 199,499                               & \cellcolor{blue!25}76\%          & 14\%            & 10\%            & 9                                                                                    \\ \hline

\end{tabular}
\caption{General user demographic per clearweb forum}
\label{tab:user_profiles}
\end{table}

\begin{table}[h!]
\scriptsize
\centering
\begin{tabular}{|c|c|c|c|c|c|}
\hline
\multirow{2}{*}{\textbf{Top 5 age groups}} & \multirow{2}{*}{\textbf{{[}20, 30)}} & \multirow{2}{*}{\textbf{{[}30, 40)}} & \multirow{2}{*}{\textbf{{[}40, 50)}} & \multirow{2}{*}{\textbf{{[}50, 60)}} & \textbf{{[}60, 70) in GR}    \\ \cline{6-6} 
                                           &                                      &                                      &                                      &                                      & \textbf{\textless{}= 20 in PH} \\ \hline
\textbf{GR}                         & 24.8\%                               & 23.4\%                               & 22.2\%                               & 11.9\%                               & 5.8\%                               \\ \hline
\textbf{PH}                           & 44.2\%                               & 30\%                                 & 12\%                                 & 6\%                                  & 4.4\%                               \\ \hline
\end{tabular}
\caption{Top 5 age groups for Goodreads and Pornhub}
\label{tab:age_group}
\end{table}

\begin{table}[ht!]
\scriptsize
\centering
\begin{tabular}{|c|c|c|c|c|c|}
\hline
\multirow{2}{*}{\textbf{}} & \multirow{2}{*}{\textbf{Top-1}} & \multirow{2}{*}{\textbf{Top-2}} & \multirow{2}{*}{\textbf{Top-3}} & \multirow{2}{*}{\textbf{Top-4}} & \multirow{2}{*}{\textbf{Top-5}} \\
                           &                                 &                                 &                                 &                                 &                                 \\ \hline
\textbf{GR}         & U.S. (59\%)            & U.K. (6.8\%)          & Canada (4\%)                    & India (2.7\%)                   & Italy (2.6\%)                   \\ \hline
\textbf{PH}           & U.S. (52.3\%)          & U.K.(7.3\%)          & Canada (4.7\%)                  & Germany (3.3\%)                 & France (2.5\%)                  \\ \hline
\end{tabular}
\caption{Top 5 countries per clearweb forum}
\label{tab:top_countries}
\end{table}

\subsubsection*{Experiments and Results}

For these experiments, we treated the additional datasets like the test partition of the PAN dataset as described earlier, performing the same preprocessing, feature extraction and machine learning techniques. 

As was shown in the previous section, the two additional datasets used for these experiments show a highly divergent data gender distribution compared to the PAN dataset, which was almost completely balanced. With regard to age group distribution, the PH dataset shows a similar distribution over our previously used categories, while the GR dataset shows a more balanced distribution over the 3 youngest groups. As these additional datasets were collected from platforms covering highly divergent topics compared to the content discussed in the PAN dataset, it was expected that the performance would decrease. Hence, our key research question was whether our models could still outperform the random baselines when applied in a cross domain set-up.  

As can be seen in Table \ref{tab:cross_age}, the performance of our message-based and user level models for age group identification dropped significantly when applied to both the PH and the GR datasets, resulting in random-like behaviour. This can be explained by the models overfitting on topic-specific age-related differences in language use expressed in the PAN dataset, but also potentially by the difference in time of collection, especially given the speed in which new linguistic varieties emerge in online communities. These findings highlight the need of platform-specific ground truth data for such analysis.

Contrary to our age group identification experiments, the gender detection models showed a higher robustness within our cross domain set-up (see Table \ref{tab:cross_gender}). Interestingly, the user level approach yielded a considerably higher f-score (69.0\%) than the post level classifier (56.4\%) when applied to the PH dataset, but the post level approach outperformed the user level classifier on the GR dataset (72.2\% vs. 62.1\%). Although this performance might not be sufficiently accurate to label a DNM user as either male or female in the context of a cybercrime investigation, the gender models could allow for a more systematic estimation of the gender distributions in Darknet Markets than the self-reporting surveys described in prior work. In the next section, we discuss our final quantitative analysis of the DNM dataset.

\begin{table}[h]
\centering
\label{tab:cross_age}
\caption{Micro f-scores and Accuracy for age group identification using the post level approach (PL) and the user level approach (UL).}
\begin{tabular}{lllll}
\hline
\multicolumn{1}{|l|}{\multirow{2}{*}{\textbf{Dataset}}} & \multicolumn{2}{c|}{\textbf{PL Class.}}                           & \multicolumn{2}{c|}{\textbf{UL Class.}}                           \\ \cline{2-5} 
\multicolumn{1}{|l|}{}                         & \multicolumn{1}{l|}{\textbf{Micro F}} & \multicolumn{1}{l|}{\textbf{Acc.}} & \multicolumn{1}{l|}{\textbf{Micro F}} & \multicolumn{1}{l|}{\textbf{Acc.}} \\ \hline
\multicolumn{1}{|l|}{Pornhub}                  & \multicolumn{1}{l|}{25.7}    & \multicolumn{1}{l|}{29.2} & \multicolumn{1}{l|}{21.6}    & \multicolumn{1}{l|}{28.4} \\ \hline
\multicolumn{1}{|l|}{Goodreads}                & \multicolumn{1}{l|}{23.1}        & \multicolumn{1}{l|}{25.3}     & \multicolumn{1}{l|}{25.1}        & \multicolumn{1}{l|}{26.0}     \\ \hline
\end{tabular}
\end{table}

\begin{table}[h]
\centering
\label{tab:cross_gender}
\caption{Micro f-scores and Accuracy for gender detection using the post level approach (PL) and the user level approach (UL).}
\begin{tabular}{lllll}
\hline
\multicolumn{1}{|l|}{\multirow{2}{*}{\textbf{Dataset}}} & \multicolumn{2}{c|}{\textbf{PL Class.}}                           & \multicolumn{2}{c|}{\textbf{UL Class.}}                           \\ \cline{2-5} 
\multicolumn{1}{|l|}{}                         & \multicolumn{1}{l|}{\textbf{Micro F}} & \multicolumn{1}{l|}{\textbf{Acc.}} & \multicolumn{1}{l|}{\textbf{Micro F}} & \multicolumn{1}{l|}{\textbf{Acc.}} \\ \hline
\multicolumn{1}{|l|}{Pornhub}                  & \multicolumn{1}{l|}{56.4}    & \multicolumn{1}{l|}{49.7} & \multicolumn{1}{l|}{69.0}    & \multicolumn{1}{l|}{65.0} \\ \hline
\multicolumn{1}{|l|}{Goodreads}                & \multicolumn{1}{l|}{72.2}        & \multicolumn{1}{l|}{72.3}     & \multicolumn{1}{l|}{62.1}        & \multicolumn{1}{l|}{62.0}     \\ \hline
\end{tabular}
\end{table}

\subsection*{DNM Analysis}
\label{sec:DNM_analysis}

\subsubsection*{Dataset}

For this analysis, we make use of over 2.5 million posts drawn from over 150,000
users from 35 cybercriminal communities, drawn from the DNM Corpus: a large dataset collected
between 2013 and 2015~\cite{dnmArchives}. In particular, we targeted discussion
fora within this collection, which acted as support areas for underground
marketplaces dealing in a number of different illicit goods.
Table~\ref{tab:comms} gives a breakdown of the data available for each
community. Communities ranged from successfully established markets with
thousands of users (though not all were always active posters) to small sites
that never moved beyond a handful of initial users.

\begin{table}[h!]
\centering
\scriptsize
\caption{Breakdown of the communities targeted for profiling in our case study}
\begin{tabular}{l  r  r }
\hline
\textbf{Community} & \textbf{Posts} & \textbf{Users} \\
\hline
Silk Road 2 & 882,418  & 26,163 \\
Silk Road & 846,077  & 52,383 \\
Evolution & 509,225  & 33,743 \\
Abraxas & 276,300  & 1,607 \\
Agora & 84,914  & 6,153 \\
BlackMarketReloaded & 80,467  & 7,006 \\
Nucleus & 65,175  & 9,478 \\
TheHub & 58,642  & 7,337 \\
Pandora & 49,023  & 8,729 \\
BlackBank & 32,817  & 2,381 \\
TheMajesticGarden & 26,121  & 1,858 \\
Utopia & 14,458  & 4,392 \\
Diabolus & 11,456  & 2,151 \\
Kingdom & 10,285  & 856 \\
ProjectBlackFlag & 6,131  & 330 \\
CannabisRoad2 & 5,842  & 2,139 \\
CannabisRoad3 & 4,905  & 1,903 \\
Bungee54 & 3,325  & 1,510  \\
Panacea & 2,241  & 520 \\
TorBazaar & 2,205  & 902 \\
TheRealDeal & 1,049  & 115 \\
Hydra & 937  & 276 \\
Kiss & 933  & 145 \\
Andromeda & 894  & 1,601 \\
OutlawMarket & 689  & 2,007 \\
Revolver & 660  & 85 \\
TorEscrow & 490  & 294 \\
DarkBay & 332  & 484 \\
Dogeroad & 300  & 118 \\
DarknetHeroes & 190  & 793 \\
Havana & 181  & 77 \\
Tom & 144  & 4,120 \\
GreyRoad & 43  & 24 \\
Tortuga & 37  & 7 \\
MrNiceGuy & 25  & 6 \\
\hline
\end{tabular}
\label{tab:comms}
\end{table}

The raw data provided in~\cite{dnmArchives} captures fora as scraped at several
semi-regular intervals by the dataset authors. This leads to heavy redundancy
within the data, as threads may be captured at multiple times. However, this
redundancy is also useful, as it helps to guard against intermittent faults in
the crawling process. Our approach to parsing the data takes a
\emph{latest-version-first} view -- of all pages captured within the crawling
process, we treat as canonical the most recent version, only parsing older pages
where they were not captured in later scrapes. We note that capturing pages from
older scrapes is an important step in handling this data, as many thousands of
threads and user profile pages are not present at all in the most recent scrapes
of each forum.  Differences could be attributed to crawling failures in later
scrapes, incomplete coverage as part of the crawling processes, or to
administrator action in taking down or hiding discussion threads over time.

Parsing of the data proceeded in two stages within the scrape history of each
community. First, user profile pages were processed to build up a dataset of
users and associated information from their profile pages (e.g., PGP public
keys, membership status).  Next, discussion thread pages were parsed in order to
associate posts (including textual content and metadata such as posting time,
subforum, etc.) with the user that authored them. Where quotations of other
users could be identified within the text of a user's post, these quotations
were separated from the authored text, to avoid contamination of profiling
analysis.  It sometimes occurred that user profile pages were not captured in
the scrapes due to sites protecting access to those pages, or where users were
observed posting for whom no profile page had been seen (either due to people
using guest accounts, or due to incomplete coverage of profile pages in the
crawls).  In these cases, new user entries were created on the fly during the
second stage of parsing, using such metadata as was available about the author
account from the post metadata. 

\subsubsection*{Experiments and Results}

Based on our results described in the previous section, we applied the post level age group identification model and the user level gender detection model on each DNM forum. All users that produced at least 1 message were included in the analysis. 

The results of our analysis are shown in Table \ref{tab:DNM_analysis}. With regard to age group identification (keeping the caveat in mind that we did not have access to any ground truth data about DNM user demographics and our models showed a considerable drop in performance for this task when applied across different online domains), our results seem to confirm prior work in this area and are in line with the findings of our qualitative analysis: the majority of users were labelled as being between 18 and 24 (58.9\%), with the second largest group of users labelled as 25 to 34 years old (21.1\%). 
However, the results for gender detection differ remarkably from previous studies relying on self-reporting interviews with cyber offenders, suggesting that women may be more engaged in Darknet Markets than traditionally considered. However, this hypothesis requires further research that includes verification of our models against writing prints of arrested cyber offenders to be confirmed.

\begin{table}[h!]
  \centering
  \small
  \caption{Results for the DNM dataset per age and gender group based on the age group and gender detection models (no ground truth available).}
    \begin{tabular}{|lcccc|cc|}
    \hline
    \textbf{Forum} & \textbf{18-24} & \textbf{25-34} & \textbf{35-49} & \textbf{50-XX} & \textbf{Female} & \textbf{Male} \\
    \hline
    abraxas & 282   & 204   & 74    & 21    & 232   & 349 \\
    agora & 3,277 & 1,885 & 691   & 295   & 2,933 & 3,215 \\
    andromeda & 118   & 101   & 39    & 12    & 113   & 157 \\
    blackbank & 1,536 & 497   & 244   & 102   & 1,250 & 1,129 \\
    bmr   & 3,917 & 1,956 & 759   & 372   & 3,069 & 3,935 \\
    bungee54 & 240   & 66    & 48    & 16    & 191   & 179 \\
    cannabisroad2 & 941   & 344   & 171   & 71    & 852   & 675 \\
    cannabisroad3 & 936   & 353   & 164   & 78    & 860   & 671 \\
    crackingarena & 6,063 & 1,269 & 2,309 & 2,317 & 6,759 & 5,199 \\
    crackingfire & 6,790 & 2,112 & 3,388 & 2,191 & 6,839 & 7,642 \\
    darkbay & 29    & 23    & 18    & 1     & 37    & 34 \\
    darknetheroes & 21    & 15    & 9     & 1     & 20    & 26 \\
    diabolus & 358   & 159   & 64    & 27    & 228   & 380 \\
    dogeroad & 33    & 15    & 7     & 4     & 31    & 28 \\
    evolution & 13,178 & 4,647 & 2,809 & 1,304 & 10,301 & 11,637 \\
    greyroad & 7     & 3     & 2     &   0    & 6     & 6 \\
    hackhound & 350   & 230   & 146   & 65    & 325   & 466 \\
    havana & 11    & 14    & 6     & 5     & 18    & 18 \\
    hydra & 125   & 92    & 42    & 17    & 119   & 157 \\
    kingdom & 304   & 108   & 74    & 36    & 220   & 302 \\
    kiss  & 56    & 36    & 20    & 8     & 50    & 70 \\
    mrniceguy & 2     & 2     & 2     &    0   & 1     & 5 \\
    nucleus & 2,694 & 1,075 & 490   & 270   & 2,113 & 2,416 \\
    outlawmarket & 84    & 68    & 23    & 11    & 91    & 95 \\
    panacea & 92    & 29    & 37    & 7     & 86    & 79 \\
    pandora & 2,803 & 1,659 & 455   & 243   & 2,252 & 2,908 \\
    pbf   & 167   & 81    & 46    & 25    & 127   & 192 \\
    revolver & 42    & 20    & 12    & 6     & 46    & 34 \\
    silkroad & 25,461 & 7,437 & 4,541 & 1,839 & 19,194 & 20,084 \\
    silkroad2 & 14,973 & 5,025 & 2,163 & 963   & 10,232 & 12,892 \\
    thehub & 2,230 & 1,369 & 496   & 222   & 1,765 & 2,552 \\
    themajesticgarden & 1,079 & 489   & 182   & 85    & 868   & 967 \\
    therealdeal & 47    & 36    & 17    & 6     & 55    & 51 \\
    tom   & 20    & 20    & 10    & 1     & 28    & 23 \\
    torbazaar & 102   & 55    & 29    & 10    & 87    & 109 \\
    torescrow & 47    & 28    & 13    & 3     & 26    & 65 \\
    tortuga & 8     & 2     & 1     & 1     & 8     & 4 \\
    utopia & 733   & 303   & 143   & 75    & 564   & 690 \\
    webkill & 769   & 372   & 158   & 46    & 715   & 630 \\
    \hline
    \textbf{Total (\%)} & \textbf{58.9} & \textbf{21.1} & \textbf{13.0} & \textbf{7.0} & \textbf{47.6} & \textbf{52.4} \\
    \hline
    \end{tabular}
  \label{tab:DNM_analysis}
\end{table}

\begin{figure}[!ht]
  \centering
  \includegraphics[width=\textwidth]{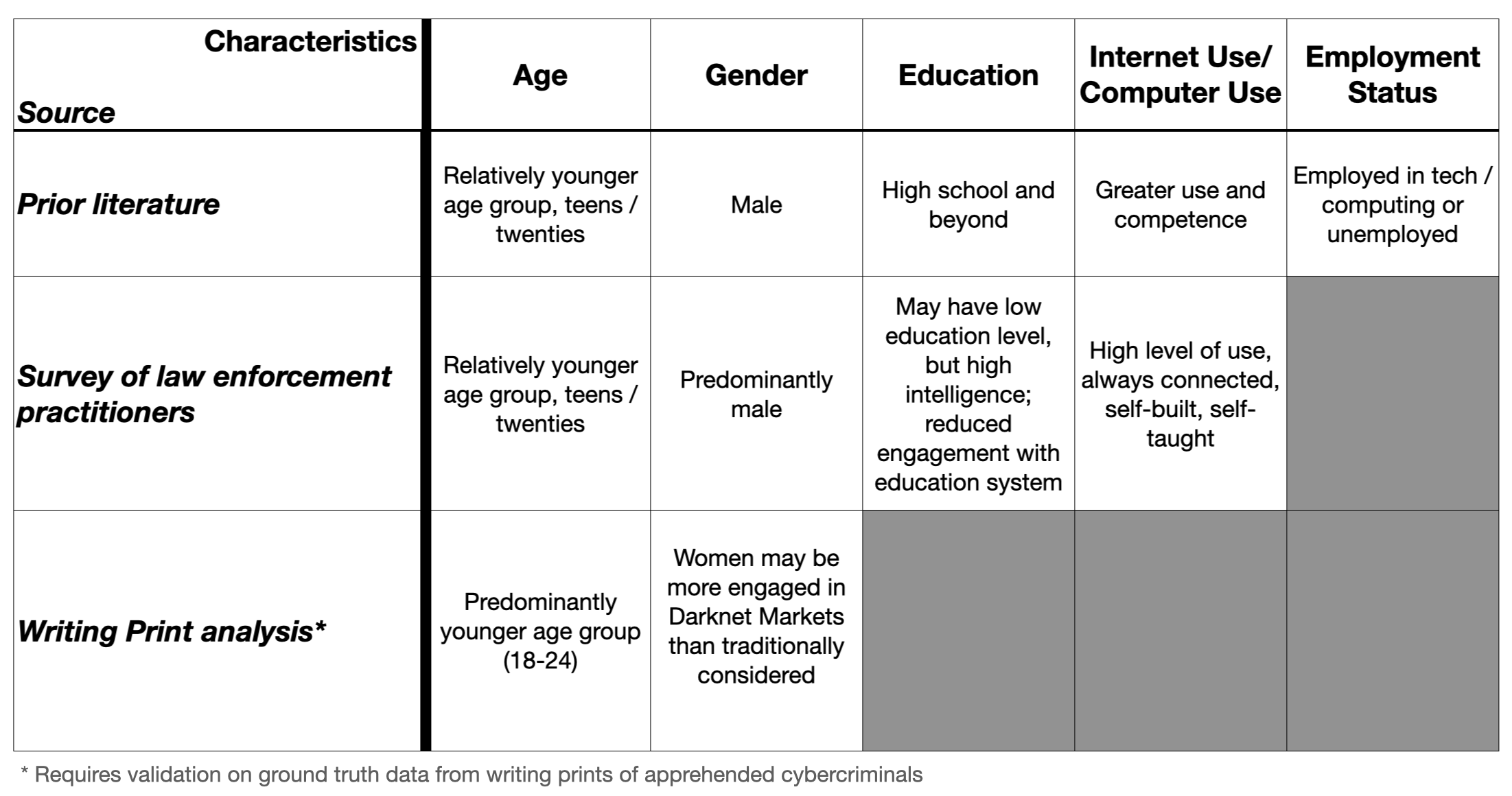}
  \caption{Contrasting characteristics identified through the three sources}
  \label{fig:amoc_characteristics}
\end{figure}

Despite the challenges of the Natural Language Processing experiments described in this Appendix, combining all aspects of the work described previously with the findings from this study did enable a novel perspective on cybercriminal motivations and characteristics, resulting in a useful framework for developing appropriate socio-technical interventions (e.g., diverting early stage offenders to more positive outlets for their skills). To conclude our work, Figure~\ref{fig:amoc_characteristics} provides an overview of the characteristics and motivations highlighted by the literature and contrasts these with the characteristics emerging from our qualitative analysis of contemporary experiences and perspectives of practitioners working within the cybercrime field\footnote{Based on quantitative and qualitative analysis of results from a survey of 16 practitioners}, and the results from the quantitative, text mining analyses presented in this study.

\subsection*{Conclusion}

To extend our qualitative analysis discussed above, we reviewed a range of Natural Language Processing and Text Categorisation techniques that could be deployed on online communications in Darknet marketplaces. Our aim was to remove the observer-observed effect, which would allow us to explore cybercriminal interactions without having to rely on the self-reporting of individual users. While recent advances in these fields have shown promising results with regard to detecting demographic characteristics, such as age group and gender, in several text genres by automatically analysing the variation of an author's linguistic features, applying such techniques on underground cybercriminal communications differs from more general applications in that its defining characteristics are both domain and process dependent. This gives rise to a number of challenges of which contemporary research has only scratched the surface. More specifically, a text mining approach applied on online communications typically has no control over the dataset size -- the number of available communications will vary across users. Hence, an automated system has to be robust towards limited data availability. Additionally, the quality of the data cannot be guaranteed. As a result, the approach needs to be tolerant to a certain degree of linguistic noise (for example, abbreviations, non-standard language use, spelling variations and errors, non-standard use of punctuation). Finally, in the context of cybercriminal fora, it has to be robust towards deceptive or \textit{adversarial} behaviour, i.e. offenders who attempt to hide their identity and criminal intentions (\textit{obfuscation}) or who assume a false digital persona (\textit{imitation}), potentially using coded language. 

Despite the challenging characteristics of this text genre for natural language processing, the present study showed that it is feasible to improve upon random baseline performance for both age group and gender classification when training on highly sparse, skewed datasets of on average 11.1 tokens per message, which also contain linguistically noisy text samples. Moreover, our message-based approach, in which we aggregate predictions on the post level to the user level using an ensemble learning method, outperformed the traditional user-based approach in which predictions are rendered directly on the user level for both age group identification and gender detection. This seems to be in line with our hypothesis that users' ``writing  prints''  might  unwittingly flow through in parts of their communications and that such clues can be identified more accurately when training on smaller text samples. These findings also seem to support and the human stylome hypothesis previously described. 

Applying our models across different online social media datasets (with a divergent data distribution over the different categories to be detected and completely divergent topics discussed), resulted in a considerable drop of the performance for age group identification, highlighting the need for up-to-date and domain-specific ground truth data to maintain a useful performance for this difficult, non binary, highly imbalanced classification task.

However, with regard to gender detection, our models were still able to significantly outperform the random baseline performance for both additional online platforms. Although this performance would not be sufficiently accurate to label a DNM user as either male or female with great confidence in a cybercrime investigation, the model does allow for a more systematic estimation of the gender distributions in Darknet Markets than the self-reporting surveys described in prior work. 

Based on our findings, we performed a final analysis of the key DNM dataset available to the AMoC team. Keeping the caveat in mind that we did not have access to any ground truth data about DNM user demographics and our models showed a considerable drop in performance for age group identification when applied across different online domains, our results seemed to confirm previous findings in literature and our qualitative analysis: the majority of users were labelled as being between 18 and 24 (58.9\%), with the second largest group of users labelled as 25 to 34 years old (21.1\%). However, the results for gender detection differed remarkably from previous studies relying on self-reporting interviews with cyber offenders, which could suggest that women may be more engaged in Darknet Markets than traditionally considered. This finding is reflected in recent work by \cite{fleetwood2020gendering} on gendering research on online illegal drug markets, who criticise the assumptions that illegal drug selling is essentially a male dominated activity and that the peripheral role of women in illegal drug selling is likely to be reproduced online. Instead, they argue that the relative anonymity afforded by Darknet Markets might hold a particular appeal for women, both as buyers and vendors of illegal drugs: selling drugs might be perceived as a way for women to achieve financial stability, independence or a sense of empowerment. Furthermore, the authors of \cite{koenraadt2018internet} found that women represented up to a third of online lifestyle drug purchasers (e.g., weight loss drugs, painkillers and sedatives). Hence, such drug advertisements could be targeting women both on clearnet and in Darknet Markets. However, these hypotheses require further research that includes verification of our models against writing prints of arrested cyber offenders to be confirmed.

\bibliographystyle{IEEEtran}
\bibliography{bibliography}

\begin{thebibliography}{10}
\providecommand{\url}[1]{#1}
\csname url@samestyle\endcsname
\providecommand{\newblock}{\relax}
\providecommand{\bibinfo}[2]{#2}
\providecommand{\BIBentrySTDinterwordspacing}{\spaceskip=0pt\relax}
\providecommand{\BIBentryALTinterwordstretchfactor}{4}
\providecommand{\BIBentryALTinterwordspacing}{\spaceskip=\fontdimen2\font plus
\BIBentryALTinterwordstretchfactor\fontdimen3\font minus
  \fontdimen4\font\relax}
\providecommand{\BIBforeignlanguage}[2]{{%
\expandafter\ifx\csname l@#1\endcsname\relax
\typeout{** WARNING: IEEEtran.bst: No hyphenation pattern has been}%
\typeout{** loaded for the language `#1'. Using the pattern for}%
\typeout{** the default language instead.}%
\else
\language=\csname l@#1\endcsname
\fi
#2}}
\providecommand{\BIBdecl}{\relax}
\BIBdecl

\bibitem{brennan2012adversarial}
M.~Brennan, S.~Afroz, and R.~Greenstadt, ``Adversarial stylometry:
  Circumventing authorship recognition to preserve privacy and anonymity,''
  \emph{ACM Transactions on Information and System Security (TISSEC)}, vol.~15,
  no.~3, p.~12, 2012.

\bibitem{brennan2009practical}
M.~R. Brennan and R.~Greenstadt, ``Practical attacks against authorship
  recognition techniques.'' in \emph{IAAI}, 2009.

\bibitem{rangel2015overview}
F.~Rangel, P.~Rosso, M.~Potthast, B.~Stein, and W.~Daelemans, ``Overview of the
  3rd author profiling task at {PAN} 2015,'' in \emph{CLEF}.\hskip 1em plus
  0.5em minus 0.4em\relax sn, 2015, p. 2015.

\bibitem{rangel2013overview}
F.~Rangel, P.~Rosso, M.~Koppel, E.~Stamatatos, and G.~Inches, ``Overview of the
  author profiling task at {PAN} 2013,'' in \emph{CLEF Conference on
  Multilingual and Multimodal Information Access Evaluation}.\hskip 1em plus
  0.5em minus 0.4em\relax CELCT, 2013, pp. 352--365.

\bibitem{rangel2014overview}
F.~Rangel, P.~Rosso, I.~Chugur, M.~Potthast, M.~Trenkmann, B.~Stein,
  B.~Verhoeven, and W.~Daelemans, ``Overview of the 2nd author profiling task
  at {PAN} 2014,'' in \emph{CLEF 2014 Evaluation Labs and Workshop Working
  Notes Papers, Sheffield, UK, 2014}, 2014, pp. 1--30.

\bibitem{sarawgi2011gender}
R.~Sarawgi, K.~Gajulapalli, and Y.~Choi, ``Gender attribution: tracing
  stylometric evidence beyond topic and genre,'' in \emph{Proceedings of the
  Fifteenth Conference on Computational Natural Language Learning}.\hskip 1em
  plus 0.5em minus 0.4em\relax Association for Computational Linguistics, 2011,
  pp. 78--86.

\bibitem{peersman2018detecting}
C.~Peersman, ``Detecting deceptive behaviour in the wild: text mining for
  online child protection in the presence of noisy and adversarial social media
  communications,'' Ph.D. dissertation, Lancaster University, 2018.

\bibitem{luyckx2011scalability}
K.~Luyckx, \emph{Scalability issues in authorship attribution}.\hskip 1em plus
  0.5em minus 0.4em\relax ASP/VUBPRESS/UPA, 2011.

\bibitem{crystal2001language}
D.~Crystal, ``Language and the {I}nternet,'' \emph{Cambridge, CUP}, 2001.

\bibitem{pedregosa2011scikit}
F.~Pedregosa, G.~Varoquaux, A.~Gramfort, V.~Michel, B.~Thirion, O.~Grisel,
  M.~Blondel, P.~Prettenhofer, R.~Weiss, V.~Dubourg \emph{et~al.},
  ``Scikit-learn: Machine learning in python,'' \emph{the Journal of machine
  Learning research}, vol.~12, pp. 2825--2830, 2011.

\bibitem{chollet2015keras}
F.~Chollet \emph{et~al.}, ``Keras,'' \url{https://github.com/fchollet/keras},
  2015.

\bibitem{sebastiani2002machine}
F.~Sebastiani, ``Machine learning in automated text categorization,'' \emph{ACM
  computing surveys (CSUR)}, vol.~34, no.~1, pp. 1--47, 2002.

\bibitem{seiffert2014empirical}
C.~Seiffert, T.~M. Khoshgoftaar, J.~Van~Hulse, and A.~Folleco, ``An empirical
  study of the classification performance of learners on imbalanced and noisy
  software quality data,'' \emph{Information Sciences}, vol. 259, pp. 571--595,
  2014.

\bibitem{scikitman2}
{scikit-learn}, ``\BIBforeignlanguage{English}{Scikit-learn user guide},''
  \url{https://scikit-learn.org/stable/}.

\bibitem{kotsiantis2007supervised}
S.~B. Kotsiantis, I.~Zaharakis, and P.~Pintelas, ``Supervised machine learning:
  A review of classification techniques,'' \emph{Emerging artificial
  intelligence applications in computer engineering}, vol. 160, pp. 3--24,
  2007.

\bibitem{breiman1984olshen}
F.~Breiman, ``Olshen, and stone,'' \emph{Classification and Regression trees},
  1984.

\bibitem{crammer2006online}
K.~Crammer, O.~Dekel, J.~Keshet, S.~Shalev-Shwartz, and Y.~Singer, ``Online
  passive aggressive algorithms,'' 2006.

\bibitem{kingma2014adam}
D.~P. Kingma and J.~Ba, ``Adam: A method for stochastic optimization,''
  \emph{arXiv preprint arXiv:1412.6980}, 2014.

\bibitem{heck2000robustness}
L.~P. Heck, Y.~Konig, M.~K. S{\"o}nmez, and M.~Weintraub, ``Robustness to
  telephone handset distortion in speaker recognition by discriminative feature
  design,'' \emph{Speech Communication}, vol.~31, no. 2-3, pp. 181--192, 2000.

\bibitem{nowak2017lstm}
J.~Nowak, A.~Taspinar, and R.~Scherer, ``Lstm recurrent neural networks for
  short text and sentiment classification,'' in \emph{International Conference
  on Artificial Intelligence and Soft Computing}.\hskip 1em plus 0.5em minus
  0.4em\relax Springer, 2017, pp. 553--562.

\bibitem{peersman2016effects}
C.~Peersman, W.~Daelemans, R.~Vandekerckhove, B.~Vandekerckhove, and
  L.~Van~Vaerenbergh, ``The effects of age, gender and region on non-standard
  linguistic variation in online social networks,'' \emph{arXiv preprint
  arXiv:1601.02431}, 2016.

\bibitem{tagliamonte2012variationist}
S.~Tagliamonte, \emph{Variationist sociolinguistics: Change, observation,
  interpretation}.\hskip 1em plus 0.5em minus 0.4em\relax John Wiley \& Sons,
  2012, vol.~40.

\bibitem{dnmArchives}
\BIBentryALTinterwordspacing
G.~Branwen, N.~Christin, D.~Décary-Hétu, R.~M. Andersen, StExo,
  E.~Presidente, Anonymous, D.~Lau, D.~K. Sohhlz, V.~Cakic, V.~Buskirk, Whom,
  M.~McKenna, and S.~Goode, ``Dark net market archives, 2011-2015,''
  \url{https://www.gwern.net/DNM-archives}, July 2015, accessed: 209-08-01.
  [Online]. Available: \url{https://www.gwern.net/DNM-archives}
\BIBentrySTDinterwordspacing

\bibitem{fleetwood2020gendering}
J.~Fleetwood, J.~Aldridge, and C.~Chatwin, ``Gendering research on online
  illegal drug markets,'' \emph{Addiction Research \& Theory}, vol.~28, no.~6,
  pp. 457--466, 2020.

\bibitem{koenraadt2018internet}
R.~Koenraadt and K.~van~de Ven, ``The internet and lifestyle drugs: an analysis
  of demographic characteristics, methods, and motives of online purchasers of
  illicit lifestyle drugs in the netherlands,'' \emph{Drugs: Education,
  Prevention and Policy}, vol.~25, no.~4, pp. 345--355, 2018.

\end{thebibliography}

\end{document}